# Fermilab PIP-II CDS and CM Cryogenic Controls System

**Pratik Patel, Ram Dhuley, Ahmed Faraj, Alexander Martinez, Vrushank Patel, William Soyars, Sungwoon Yoon**

Fermi National Accelerator Laboratory, Batavia, Illinois, United States

E-mail: pratik@fnal.gov

**Abstract.** We present an overview of the Cryogenic Electrical & Controls System for Fermilab's next generation particle accelerator Proton Improvement Plan II (PIP-II). The Controls System includes the cryogenic instrumentation and controls of PIP-II Cryogenics Distribution System (CDS) and Cryomodules (CMs). Its design is based on Siemens PCS7 Controls System with 26 Remote IO Rittal Cabinets and 48 Relay Racks that house for temperature readouts, control valve positioners, liquid helium level readouts, controllers for heater power supplies, etc. We describe the full architecture of the controls system, the choice of instrumentation considering certain design constraints such as radiation tolerance in the accelerator tunnel, and the EPICS communications protocol that is used as a SCADA system communicating to S7 Controllers *via* OPC UA.

## 1. Introduction

The Proton Improvement Plan II, or PIP-II, project is an essential upgrade to Fermilab's accelerator complex to enable the world's most intense high-energy beam of neutrinos for the international Deep Underground Neutrino Experiment at LBNF, and a broad physics program, powering new discoveries for many decades to come. PIP-II features a brand-new, 800-million-electronvolt leading-edge superconducting radio-frequency linear accelerator that will enable the Fermilab complex to deliver more than a megawatt of beam power to LBNF, upgradeable to multimegawatt levels [1].

The PIP-II cryogenic system shown in Figure 1 consists of two major subsystems: the helium cryogenic plant (Cryoplant) and the cryogenic distribution system (CDS) which collectively delivers the required cryogenic refrigeration to the Linac Cryomodules (CM). The Cryoplant provides the required cooling capacity at three nominal temperature levels: 40 K for the high temperature thermal shields and intercepts, designated as HTTS; 4.5 K for the low temperature intercepts, designated as LTTS; and 2 K for the SRF cavities and magnets within each cryomodule. The combined 4.5K supply is divided into two streams as it enters the cryomodules, one that is directed a heat exchanger, and another directed to the LTTS.

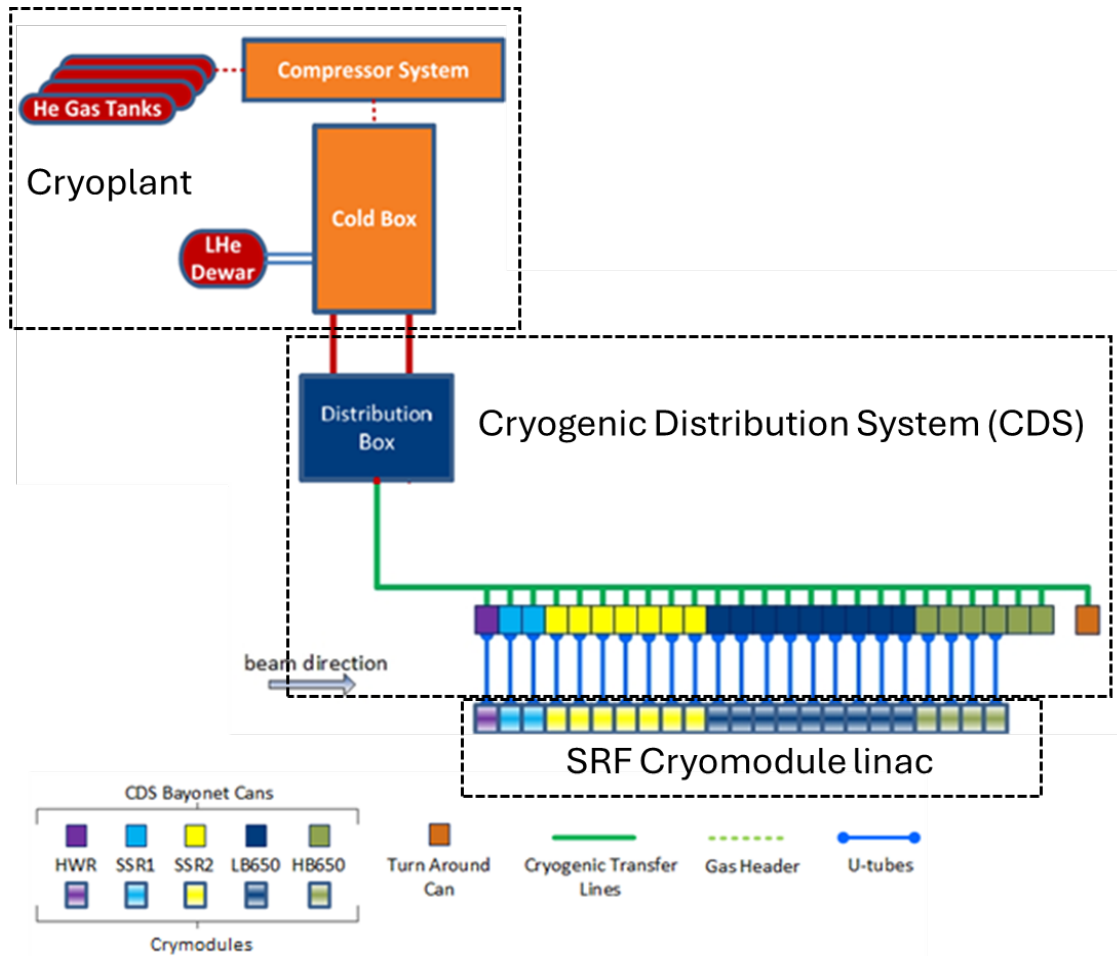

**Figure 1.** PIP-II Cryogenic System.

Depicted in Figure 2, the PIP-II Linac comprises a total of twenty-three (23) cryomodules of five (5) different types. Each cryomodule type has a different cavity design – half wave resonator (HWR), two types of single spoke resonators (SSR-1 and SSR-2), and two types of elliptical cell resonators (LB650 and HB650). The CDS system consists of a Distribution Box (DVB), ~200m of Cryogenic Transfer line comprising 25 modular bayonet boxes (Bayonet Can or BC) and a Turnaround Can (TC). Each Cryomodule has a corresponding bayonet can with the associated valves and instrumentation required to distribute cryogens from the CDS to/from the CM. There are two CDS Bayonet Cans at the far end that can provide cryogenic helium to future addition of two HB650 cryomodules. Also shown in Figure 2 are other non-cryogenic components of PIP-II and the PIP-II In-Kind contributors.

## 2. Cryogenics Controls System

*2.1 Controls description*
The PIP-II Cryogenics Controls System utilizes Siemens S7-400 class controllers. There are (2) PLC cabinets that host S7-4105H controller in redundant setup. Selection of S7410-5H allows for

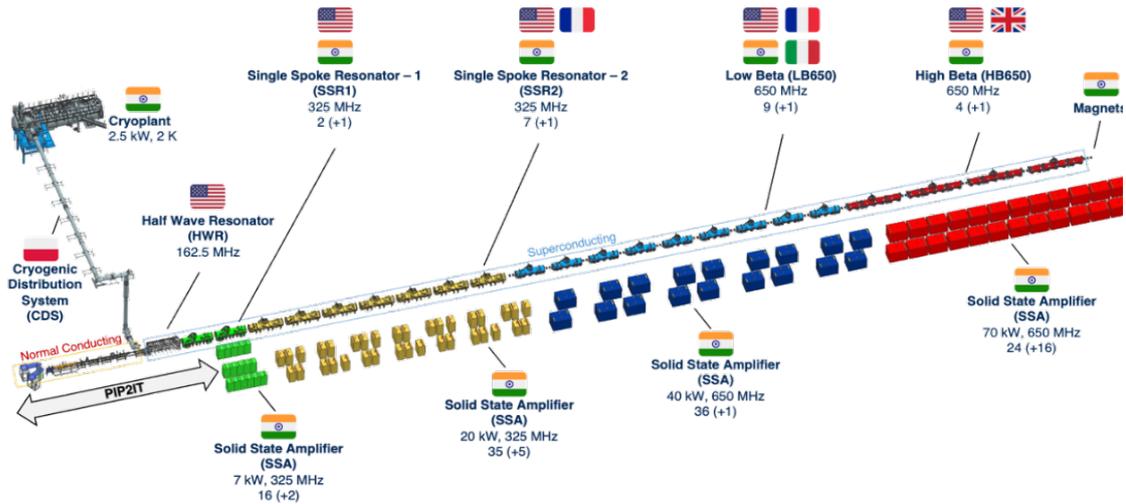

**Figure 2.** CAD models showing CDS components and the location and types of Cryomodules.

High availability & Redundant operation. Siemens PCS7 was selected because of Fermilab Cryo group's vast experience and their reliability, also PIP2IT test facility, which tests all the Cryomodules for PIP-II uses S7400 controller. Remote IO's are Siemens ET200SPHA, Remote IO monitors and controls instrumentation and valves for its respective Cryomodule and Bayonet can. There are 24x Remote IO cabinets, one for each CM+BC and one for DVB. PLC communicates to each Remote IO via redundant star Profinet IO network topology using Siemens Scalance Switches. There are redundant switches in PLC cabinets, one connected to each PLC, primary and backup one. In remote IO Cabinets there are also 2 Scalance Switches connected to each PLC, which connects to 2 redundant Profinet modules on ET200SPHA.

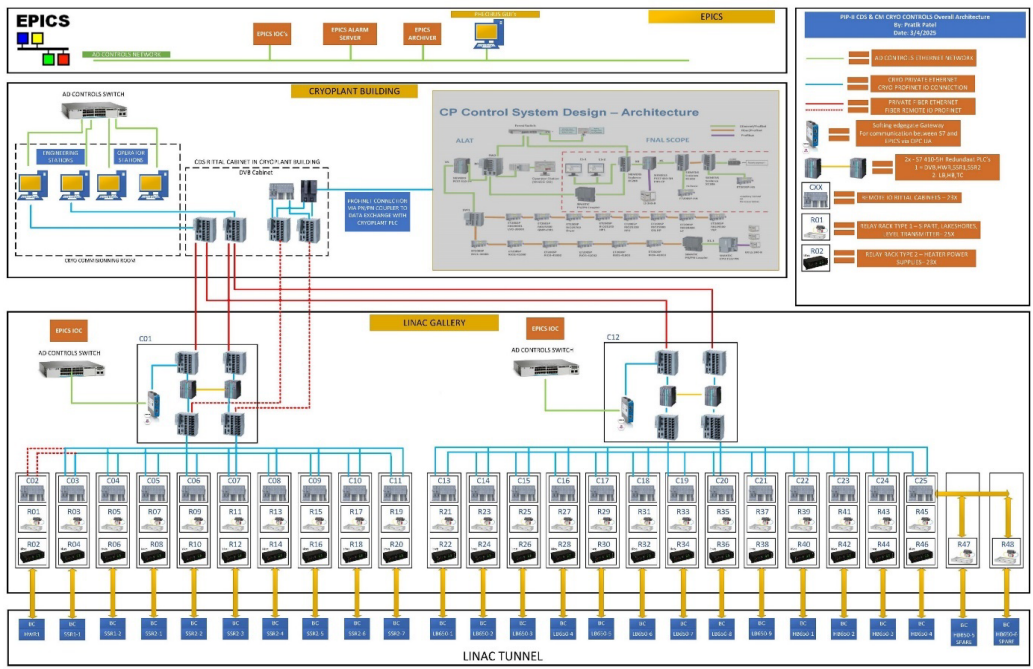

**Figure 3.** PIP-II CDS & CM Controls Overall Architecture.

Overall architecture in Figure 3 describes the PLC assignment for CDS and Cryomodules. PLC 1&2 in Cabinet 1, controls and monitors DVB, HWR, SSR1 and SSR2. PLC3&4 in Cabinet 12, controls and monitors LB650, HB650, Turnaround Can. Most connections are Ethernet RJ45, for longer distance we are using Fiber connection on Scalance.

Figure 4 shows more detailed connections witinh te architecture, For each Profinet Remote IO connection there is Profibus network. So each system has its own profibus master. Connections from Scalance in each Remote IO cabeinet goes to a Siemens Y-Switch, Y-switch is then connected to Siemens IE/PB Link HA. IE/PB Link is Master Profibus network for Remote IO. Profibus Master is then connected to Profihub switch which has 5 Profibus Channels, each channels can support 31 profibus slaves/devices. CDS PLC's communicates and shares data with Cryoplant PLC's via PN-PN coupler located in DVB cabinet .

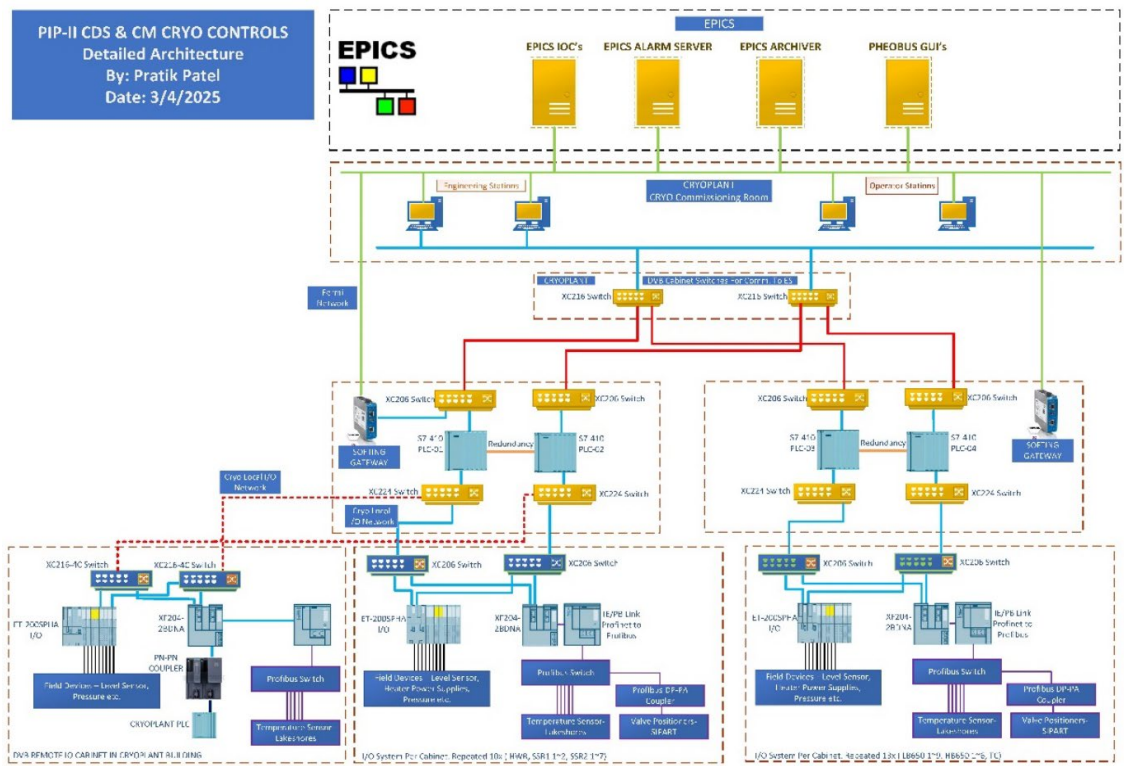

**Figure 4.** PIP-II CDS & CM Controls Detailed Architecture.

Table 1 and Table 2 shows the total instrumentation for each system, cryomodule count includes valves and instrumentation on bayonet can. In total there are 2 PLC Cabinets, 24 Remote IO Cabinets, and 48 Relay Racks. There are 2 Relay Racks for each Remote IO cabinet or per each CM. Relay Racks Type #1 has components such as SIPART Valve Positioners, Lakeshore Temperature Monitors, AMI Liquid level Transmitter, Relay Racks Type #2 has Heater Power Supplies and components. Cabinets and Relay Racks final layout in the Linac Gallery floor is designed considering Profinet and other protocols length limit: Profinet <= 100 m, Profibus <= 1200 m, and fiber <=2000 m.

Table 1. Instrumentation Count per Cryomodule Type.

| Instrumentation Quantity per CM Type | DVB x 1 | HWR x 1 | SSR1 x 2 | SSR2 x 7 | LB650 x 9 | HB650 x 4 | HB650 (Spare) x2 | TC x 1 | Totals |
|---|---|---|---|---|---|---|---|---|---|
| Platinum 100 | 0 | 28 | 30 | 20 | 14 | 20 | 0 | 0 | **434** |
| Cernox (CX-1030) | 6 | 44 | 47 | 37 | 23 | 27 | 6 | 6 | **736** |
| Pressure Transducers | 9 | 10 | 9 | 9 | 9 | 9 | 7 | 10 | **241** |
| Control Valves | 15 | 20 | 15 | 15 | 15 | 15 | 15 | 7 | **402** |
| Liquid Level | 0 | 2 | 1 | 1 | 1 | 1 | 0 | 1 | **25** |
| Flowrate transmitters | 2 | 0 | 0 | 0 | 0 | 0 | 0 | 0 | **25** |
| Heater | 0 | 5 | 14 | 10 | 6 | 8 | 0 | 1 | **190** |
| Vacuum Transducers | 0 | 1 | 1 | 1 | 1 | 1 | 0 | 0 | **23** |
| RF Interlock | 0 | 1 | 1 | 1 | 1 | 1 | 0 | 0 | **23** |
| Rupture Disc Indicator | 0 | 1 | 1 | 1 | 1 | 1 | 0 | 0 | **23** |

Table 2. I/O Count per Cryomodule Type.

| I/O Quantity per CM Type | DVB x 1 | HWR x 1 | SSR1 x 2 | SSR2 x 7 | LB650 x 9 | HB650 x 4 | HB650 (Spare) x2 | TC x 1 | Totals |
|---|---|---|---|---|---|---|---|---|---|
| AI Channel | 12 | 25 | 40 | 32 | 24 | 28 | 7 | 13 | **696** |
| AO Channel | 0 | 5 | 14 | 10 | 6 | 8 | 0 | 1 | **190** |
| DI Channel | 0 | 16 | 43 | 31 | 19 | 25 | 0 | 3 | **593** |
| DO Channel | 0 | 6 | 15 | 11 | 7 | 9 | 0 | 1 | **213** |
| LK 240-8P Channel | 6 | 72 | 77 | 57 | 37 | 47 | 0 | 6 | **1158** |
| SIPART PS2 Channel | 15 | 20 | 15 | 15 | 15 | 15 | 15 | 7 | **402** |

2.2 Instrumention and Controls

*a. Temperature measurement*

In Each cryogenic equipment will have various temperature sensors for monitoring and control. Depending on the temperature to be read there are two types of temperature sensors used,

Cernox 1030 and Platinum PT-103 (RTD). All thermometers are read using Lakeshore 240-8P temperature input modules via a four-wire measurement.

The Lakeshore 240-8P modules are connected to the PLC via Profibus DP Master network in each Remote IO on a dedicated Profinet to Profibus network.

*b. Pressure measurement*

Each pressure transmitter outputs a 4-20 mA signal that is connected on a 2-wire, 4-20mA loop to the associated I/O module. There are total 10 Pressure transducers per CM+BC. Absolute pressure transmitters used include MKS 230E and MKS 750D with measurement ranges of 0-100 torr and 0-100 psia. Gauge pressure transmitters used include Setra 206 sensors with measurement ranges from 0-500 psig. Insulating Vacuum Pressure transmitters will utilize MKS Series 925 MicroPirani Transducers with a measurement range of 1.0E-5 Torr to Atmosphere

The tunnel pressure transducers are in a radiation area, so they have no active electronics and instead have mV/V output. Dataforth Strain Gage Input Signal Conditioner is used to provide exicitation and convert mV/V output to 4-20 ma signal to PLC input.

*c. Helium liquid level measurement*

Each Cryomodule and the Turnaround Can will have a AMI 2K level probe with a 59 cm active length (to be confirmed) for measuring the liquid helium levels within the cryogenic equipment. We will utilize AMI Model 1700 transmitter box 2K configured to condition the level probe readback to a 4-20 mA signal that is than connected to the associated I/O module for readback to the PLC.

*d. Process control valves*

There are 15 pneumatic controlled valves per Cryomodule slot except HWR which has 20, 15 pneumatic valves at the DVB and 7 pneumatic valves at the Turnaround Can. The valve positioners shall utilize the specially modified Siemens SIPART PS2 originally designed for use at CERN [2]. The positioner is composed of two parts, the rack mounted microprocessor portion and the modified positioner locally mounted to each valve. The locally mounted positioner includes the piezos for supply and exhaust of the instrument air and potentiometer for reading the valve position feedback this portion of the positioner is the scope of the FNAL. See Figure 5 for simplified configuration and scope boundaries. 15 positioner channels with microprocessors occupy one rack mounted set. 2 pairs of Four shielded, twisted pair cables will run from each valve positioner to the controlling rack mounted microprocessor. Initialization is required for each positioner once compressed air is available to the valve [3].

The rack mounted positioners communicate via Profibus PA. We will utilize a Profibus DP/PA coupler to interface with the PLC Profibus DP master in each Remote IO Cabinet. The coupler Pepperl+Fuchs KFD2-BR-1.PA.1500 have been successfully used for this purpose in other projects.

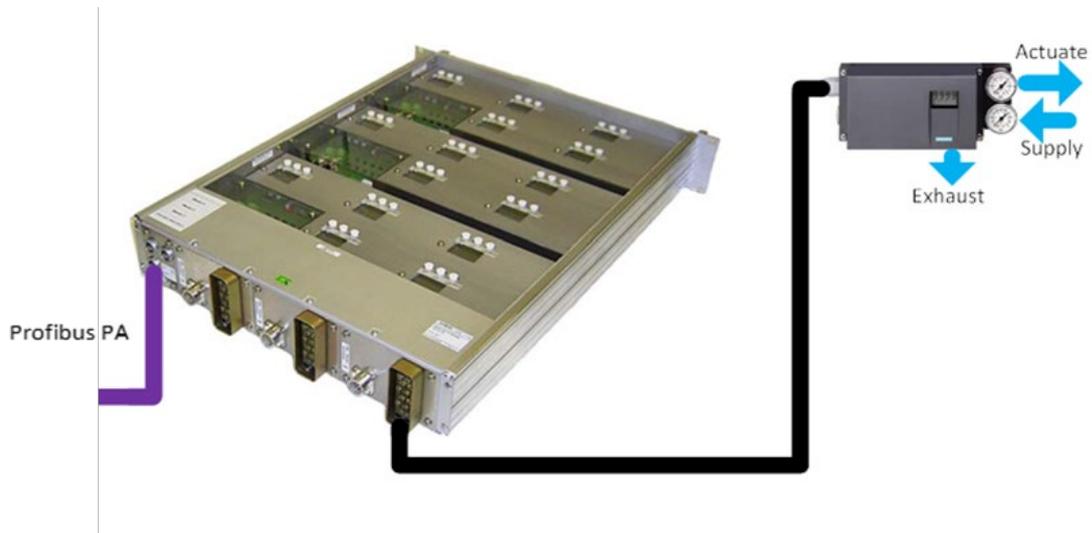

**Figure 5.** Siemens SIPART PS2 with 15 channel rack mounted chassis and remote modified positioner.

*e. Heater control*

Cryomodules contain up to four different types of heaters with various groupings/configurations. These configuration and their controls are summarized below.

Coupler Heaters:
- Dalton 10W (1/4" OD x 1" long) Model W2C010. Each cryomodule cavity has one coupler and each coupler contains 4 heaters.
- The 4 heaters per coupler will be grouped and wired in parallel and shall be powered by a single power supply

Cavity Heaters:
- Heater Minco HAP6949. For each Cryomodule, all cavity heaters will be wired in parallel and shall be powered by a single power supply.

Helium Can Heaters:
- Dalton model W4C050 (1/2" OD x 5" long), 100 Watts, 50 Volts Cartridge Heater
- The heater has four-wires (current/voltage) for powering/monitoring of heater
- Each Cryomodule has one helium can heater. Each helium can heater is protected from over temperature by liquild level interlock and operated at maximum of 50 W.

Current Lead Heaters (SSR1 and SSR2 Cryomodules):
- Dalton 10W (1/4" OD x 1" long) Model W2C010
- There is one current lead associated with each solenoid, four in SSR1 and three in SSR2

Each Heater PS is integrated with PLC using combination of external components and direct PS interface for heater control and monitoring:
- Relay – Enable/Disable Power Supply Output
- Heater PS Alarm feedback
- 4-20ma Signal from PLC Analog Out to Heater PS for Voltage Control
- Voltage Transducer for voltage feedback.
- Heater PS Current output for current feedback

*f. EPICS communication*

EPICS is selected as a SCADA system for PIP-II. EPICS IOC's have been standardized to use OPC UA protocol to commincate with all Field controllers.We will use Softing EdgeGate Gateway for data handshake between S7 Controllers and EPICS IOC using OPC UA protocol. EdgeGate is currently being commissioned at PIP-II CM Test Facility.Softing gateway can be setup via easy web interface, no software's needed. 2 independent ports allows for isolation of Fermi Network and Private PLC network.Use of gateway will prevent dependency on Windows based Computer and updates and will have dedicated hardware for communication. Gateway solution would replace existing Windows based Computers used for communication using Kepware OPCUA.

Phoebus Screens, Alarming, Datalogging development is in initial phase. PIP-II Cryoplant Mycom Compressor screens have been created.

## 3. Status and outlook

Significant progrees has been made so far towards the design and fabrication of the 24 Rittal Cabinets and 48 Relay Racks. Final Design Review for the whole Controls System was completed in 2023. All Primary Materials have been procured.

All electrical design, drawings and BOM have been completed in 2024. Local UL508A panel shop has been awarded contract to fabricate 24x Rittals and 48 Relay Racks. Fermilab procured materials have been shipped to sub-contractor and initial cabinets are being built. Cable pull database has been established and 50% completed. Programming will start in Mid June 2025, EPICS HMI Development will starts in March 2026.

**Acknowledgments**

This paper has been authored by FermiForward Discovery Group, LLC under Contract No. 89243024CSC000002 with the U.S. Department of Energy, Office of Science, Office of High Energy Physics. The authors wish to recognize the dedication and skills of the Cryogenics Department technical personnel involved in the design and testing of the cryogenic equipment.

**References**

[1]	R. Stanek, "PIP-II PROJECT OVERVIEW AND STATUS" FNAL, Batavia, IL FERMILAB-CONF-23-331-PIP2
[2]	SIPART Electropneumatic positioners SIPART PS2 with PROFIBUS PA, Siemens, 2013
[3]	Juan Casas, "Radiation Hardness of the Siemens SIPART intelligent valve positioner CERN Geneva, Switzerland, dept TE-CRG, 1211